\documentclass[12pt]{article}

\usepackage{epsfig}
\usepackage{amssymb,amsmath}
\usepackage{multirow}
\usepackage{rotating}
\usepackage{graphicx}
\usepackage{lscape}
\usepackage{hyperref}

\newcommand{\bea}{\begin{eqnarray}}
\newcommand{\eea}{\end{eqnarray}}

 \makeatletter
\def\alt{\mathrel{\mathpalette\gl@align<}}
\def\agt{\mathrel{\mathpalette\gl@align>}}
\def\gl@align#1#2{\lower.6ex\vbox{\baselineskip\z@skip\lineskip\z@
\ialign{$\m@th#1\hfil##\hfil$\crcr#2\crcr\sim\crcr}}} \makeatother

\begin{document}

\begin{flushright}
%preprint number \\
\end{flushright}

\vspace*{1.0cm}

\begin{center}
\baselineskip 20pt 
{\Large\bf 
Discrimination of Supersymmetric Grand Unified Models in Gaugino Mediation 
}
\vspace{1cm}

{\large 
Nobuchika Okada$^{a,}$\footnote{ E-mail: okadan@ua.edu} 
and 
Hieu Minh Tran$^{b,c,}$\footnote{ E-mail: hieutm-iep@mail.hut.edu.vn} 
} 
\vspace{.5cm}

{\baselineskip 20pt \it
$^a$Department of Physics and Astronomy, University of Alabama, \\ 
Tuscaloosa, Alabama 35487, USA \\
\vspace{2mm} 
$^b$Hanoi University of Science and Technology, 1 Dai Co Viet Road, Hanoi, Vietnam \\
\vspace{2mm}
$^c$Hanoi University of Science - VNU, 334 Nguyen Trai Road, Hanoi, Vietnam
}

\vspace{.5cm}

\vspace{1.5cm} {\bf Abstract}
\end{center}

We consider supersymmetric grand unified theory (GUT) 
 with the gaugino mediated supersymmetry breaking 
 and investigate a possibility to discriminate 
 different GUT models in terms of predicted sparticle mass spectra. 
Taking two example GUT models, 
 the minimal $SU(5)$ and simple $SO(10)$ models, 
 and imposing a variety of theoretical and experimental constraints, 
 we calculate sparticle masses. 
Fixing parameters of each model so as to result in 
 the same mass of neutralino 
 as the lightest supersymmetric particle (LSP), 
 giving the observed dark matter relic density, 
 we find sizable mass differences in the left-handed slepton 
 and right-handed down-type squark sectors in two models, 
 which can be a probe to discriminate the GUT models 
 realized at the GUT scale far beyond 
 the reach of collider experiments.

\thispagestyle{empty}

%\bigskip
\newpage

\addtocounter{page}{-1}

%%%%%%%%%%%%%%%%%%%%%%%%%%
%\baselineskip 36pt
% Main body
%%%%%%%%%%%%%%%%%%%%%%%%%%
\baselineskip 18pt
%%%%%%%%%%%%%%%%%%%%%%%%%%

%%%%%%%%%%%%%%%%%%%%%%%%%%
\section{Introduction}  
%%%%%%%%%%%%%%%%%%%%%%%%%%

Providing a promising solution to a long-standing problem 
 in the standard model (SM), the gauge hierarchy problem, 
 and motivated by the possibility of being tested 
 at the Large Hadron Collider (LHC) and 
 other future collider projects 
 such as the International Linear Collider (ILC), 
 supersymmetry (SUSY) has been intensively explored 
 for the last several decades. 
In addition, under the R-parity conservation, 
 the minimal supersymmetric extension of the SM (MSSM) 
 provides neutralino lightest supersymmetric particle (LSP) which is a good candidate 
 for the dark matter, a mysterious block of the universe 
 needed to explain the cosmological observation. 
Furthermore, in the MSSM, all the SM gauge couplings successfully 
 unify at the grand unified theory (GUT) scale $M_{GUT} \simeq 2 \times 10^{16}$ GeV, 
 and this fact strongly supports the GUT paradigm.

The exact SUSY requires that the SM particles and their superpartners 
 have equal masses. 
However, we have not yet observed any signal of sparticles 
 in either direct and indirect experimental searches. 
This implies that not only should SUSY be broken at some energy, 
 but also that SUSY breaking should be transmitted to the MSSM sector 
 in a clever way so as not to cause additional flavor changing neutral 
 currents and CP violations associated with supersymmetry breaking terms. 
There have been several interesting mechanisms 
 for desirable SUSY breaking and its mediations.

In this paper, we consider one of the possibilities, 
 the gaugino mediated SUSY breaking (gaugino mediation) \cite{gMSB}. 
With a simple 5D braneworld setup of this scenario, 
 the SUSY breaking is first mediated to the gaugino sector, 
 while sfermion masses and trilinear couplings 
 are negligible at the compactification scale 
 of the extra fifth dimension. 
At low energies, the sfermion masses and trilinear couplings 
 are generated through RGE runnings with the gauge interactions, 
 realizing the flavor-blind sfermion masses.  
However, the gaugino mediation in the context of the MSSM 
 predicts stau LSP, and such a stable charged particle 
 is disfavored in the cosmological point of view. 
This problem can be naturally solved 
 if the compactification scale is higher than the GUT scale 
 and a GUT is realized there \cite{spectrumgMSB}.  
The RGE runnings as the GUT play the crucial role 
 to push up stau mass, and neutralino LSP is realized 
 at the electroweak scale, which is a suitable dark matter candidate 
 as usual in SUSY models.

There are many possibilities of GUT models 
 with different unified gauge groups and representations 
 of the matter and Higgs multiplets in the groups. 
A question arising here is how we can discriminate GUT models 
 by experiments carrying out at energies 
 far below the GUT scale. 
Note that SUSY GUT models with SUSY breaking mediations 
 at or above the GUT scale leave their footprints 
 on sparticle mass spectra at low energies 
 through the RGE evolutions. 
Typical sparticle mass spectrum, once observed,  
 can be a probe of $SU(5)$ unification \cite{softprobe}.  
In a similar way, three different types of seesaw mechanism 
 for neutrino masses can be distinguished 
 at the LHC and the ILC \cite{seesaw}. 
In this paper, based on the same idea, 
 we investigate a possibility to discriminate 
 different GUT models with the gaugino mediation. 
A remarkable feature of the gaugino mediation is 
 that the model is highly predictive and 
 sparticle masses are determined by only 2 free parameters, 
 the compactification scale ($M_c$) 
 and the input gaugino mass ($M_G$) at $M_c$, 
 with a fixed $\tan \beta$ and the sign of the $\mu$-parameter.

The structure of this paper is as follows:
In Sec. 2, we briefly discuss the basic setup of 
 the gaugino mediation and introduce two examples of 
 GUT models, the minimal $SU(5)$ model and a simple $SO(10)$ model. 
In Sec. 3, we analyze the RGE evolutions of sparticle masses 
 and the trilinear couplings for the two GUT models 
 from the compactification scale to the electroweak scale, 
 and find sparticle mass spectra which are consistent 
 with a variety of theoretical and experimental constraints. 
Fixing parameters in both models to result in 
 the same neutralino LSP mass, giving the observed dark 
 matter relic abundance, we compare sparticle mass spectra. 
We find sizable sparticle mass differences which can be 
 a probe to discriminate the GUT models. 
The Section 4 is devoted for conclusions.

%%%%%%%%%%%%%%%%%%%%%%%%%%%%%%
\section{Model setup}
%%%%%%%%%%%%%%%%%%%%%%%%%%%%%%

In the gaugino mediation scenario \cite{gMSB}, 
 we introduce a 5-dimensional flat spacetime 
 in which the extra fifth dimension is compactified 
 on the $S^1/Z_2$ orbifold with a radius $r=1/M_c$. 
The SUSY breaking sector resides on a $(3+1)$-dimensional brane 
 at one orbifold fixed point, while the matter and Higgs sectors 
 are on another brane at the other orbifold fixed point. 
Since the gauge multiplet propagates in the bulk, 
 the gaugino can directly couple with the SUSY breaking sector 
 and acquires the soft mass at the tree level. 
On the other hand, due to the sequestering between two branes, 
 the matter superpartners and Higgs fields cannot directly communicate with 
 the SUSY breaking sector, hence sfermion and Higgs boson 
 soft masses and also the trilinear couplings are all zero 
 at the tree level. 
According to this structure of the gaugino mediation, 
 in actual analysis of RGE evolutions for soft parameters, 
 we set nonzero gaugino mass at the compactification scale 
 and solve RGEs from $M_c$ toward low energies. 
Soft masses of matter superpartners and Higgs fields 
 are generated via the RGE evolutions.

When the compactification scale is lower than $M_{GUT}$, 
 the detailed study on MSSM sparticle masses 
 in the gaugino mediation showed that the LSP is stau 
 in most of the parameter space \cite{spectrumgMSB}. 
Clearly, this result is disfavored in the cosmological point of view. 
However, it has been shown that this drawback can be ameliorated 
 if we assume a GUT model and $M_c > M_{GUT}$ \cite{spectrumgMSB}: 
 the RGE evolutions from $M_c$ to $M_{GUT}$ 
 push up stau mass and realize neutralino LSP. 
In other words, the grand unification is crucial 
 to realize phenomenologically viable sparticle 
 mass spectrum in the gaugino mediation. 
In order to suppress sfermion masses compared 
 to gaugino masses at the compactification scale, 
 the spatial separation between two branes 
 should not be too small; equivalently, 
 the compactification scale should not be too large. 
In the following analysis, we set the reduced Planck scale ($M_P$)
 as the upper bound on $M_c$: 
\begin{equation}
  M_c \leq M_P = 2.43 \times 10^{18} \; \rm GeV .  
\label{Planck}
\end{equation}

There have been many GUT models proposed 
 based on different unified gauge groups 
 such as $SU(5)$, $SO(10)$, and $E_6$.
In this paper, we consider two GUT models as examples, 
 namely, the minimal $SU(5)$ model 
 and a simple $SO(10)$ model \cite{so10}.

In the minimal $SU(5)$ model, 
 the matter multiplets of the $i$th generation 
 are arranged in 2 representations, 
 ${\bar{\bf{5}}}_{i}$ and ${\bf{10}}_{i}$. 
Two Higgs doublets in the MSSM are embedded in 
 the representations of ${\bar{\bf{5}}}_{H} + {\bf{5}}_{H}$, 
 while the ${\bf{24}}_{H}$ Higgs multiplet 
 plays the role of breaking the $SU(5)$ gauge symmetry to the SM one. 
The particle contents of the minimal $SU(5)$ model
 along with the Dynkin index and the quadratic Casimir 
 for corresponding multiplets are listed in Table \ref{su5}.

\begin{table}[h]
\caption{
Particle contents of the minimal $SU(5)$ GUT} \label{su5}
\begin{center}
\begin{math}
\begin{array}{c|c|c|c} 
SU(5)              & \rm Particles              & \rm Dynkin \; Index & C_{2}(\bf{R}) \\
\hline
{\bar{\bf{5}}}_{i} &  D^{c}_{i}, L_{i}          & 1/2          & 12/5 \\

{\bf{10}}_{i}      &  Q_{i},U^{c}_{i},E^{c}_{i} & 3/2          & 18/5 \\
\hline
{\bar{\bf{5}}}_{H} &  H_{d}                     & 1/2          & 12/5 \\

{\bf{5}}_{H}       &  H_{u}                     & 1/2          & 12/5 \\
\hline
{\bf{24}}_{H}      & \rm additional \; Higgs    & 5            & 5    \\
\end{array}
\end{math}
\end{center}
\end{table}

In $SO(10)$ GUT models, all the matter multiplets 
 of the $i$th generation are unified 
 into a single ${\bf{16}}_{i}$ representation. 
In a simple $SO(10)$ model investigated in \cite{so10}, 
 Higgs multiplets of the representations 
 ${\bf{10}}_{H} + {\bf{10}}'_{H} + 
 \bar{\bf{16}}_{H} + {\bf{16}}_{H} + {\bf{45}}_{H}$ are introduced. 
The up-type (down-type) Higgs doublets in the MSSM 
 are realized as a linear combination of 
 two up-type (down-type) Higgs doubles 
 in ${\bf{10}}_{H} + {\bf{10}}'_{H}$, 
 while the Higgs multiplets of 
 $\bar{\bf{16}}_{H} + {\bf{16}}_{H} + {\bf{45}}_{H}$ representations 
 work to break the $SO(10)$ gauge symmetry to the MSSM one.
Similarly to Table \ref{su5}, the particle contents of this model 
  are listed in Table \ref{so10}. 

\begin{table}[h]
\caption{
Particle contents of a simple $SO(10)$ GUT} \label{so10}
\begin{center}
\begin{math}
\begin{array}{c|c|c|c} 
SO(10)            & \rm Particles               & \rm Dynkin \; Index & C_{2}(\bf{R}) \\
\hline
{\bf{16}}_{i}     & i \rm -th \; generation         &   2          & 45/8 \\
\hline
{\bf{10}}_{H}     &  H_{u}^1, H_{d}^1               &   1          & 9/2 \\

{\bf{10}}'_{H}    &  H_{u}^2, H_{d}^2               &   1          & 9/2 \\
\hline
\bar{\bf{16}}_{H} & \multirow{3}*{additional Higgs} &   2          & 45/8 \\

{\bf{16}}_{H}     &                                 &   2          & 45/8 \\

{\bf{45}}_{H}     &                                 &   8          & 8    \\

\end{array}
\end{math}
\end{center}
\end{table}

%%%%%%%%%%%%%%%%%%%%%%%%%%%%%%%%%%%%%%%%%%
\section{Sparticle masses in two models}
%%%%%%%%%%%%%%%%%%%%%%%%%%%%%%%%%%%%%%%%%%

Now we analyze sparticle mass spectrum at low energy 
 for each GUT model. 
In the gaugino mediation, gaugino mass is 
 a unique input at the compactification scale $M_c > M_{GUT}$. 
For a given GUT model, solving the RGEs from $M_c$ to $M_{GUT}$ 
 with the gaugino mass input, we obtain a set of soft parameters 
 at the GUT scale, with which 
 we solve the MSSM RGEs for the soft parameters toward low energies. 
General 1-loop RGE formulas for the soft parameters 
 in a GUT model are given by \cite{spectrumgMSB}:
\begin{eqnarray}
 && \frac{d \alpha_U}{d t} = - \frac{b_U}{2 \pi} \alpha_U^2,  \\ 
 && \frac{d}{d t}\left( \frac{M}{\alpha_U}\right) = 0,   \\
 && \frac{d m^2}{d t} = - 2 C_2({\bf R}) \frac{\alpha_U}{\pi} M^2, \\ 
 && \frac{d A}{d t} = \left( \sum_i C_2({\bf R}_i) \right) 
   \frac{\alpha_U}{\pi} M, 
\end{eqnarray}
where $\alpha_U$ is the unified gauge coupling, 
 $b_U$ is the beta function coefficient, 
 $M$ is the running gaugino mass, 
 $m$ is the running mass of a scalar field 
 in the ${\bf R}$ representation under the GUT gauge group, 
 and $C_2$ is the quadratic Casimir.  
For the boundary conditions in the gaugino mediation scenario,
\begin{eqnarray}
 M(M_c) = M_G \neq 0, \; m^2(M_c) = 0, \; A(M_c) = 0,
\end{eqnarray}
 we can easily find the solutions: 
\begin{eqnarray}
 &&  \alpha_U(\mu)^{-1} =  \alpha_U(M_c)^{-1} 
 + \frac{b_U}{2 \pi} \ln(\mu/M_c), \\ 
 && m^2(\mu)=2 \frac{C_2({\bf R})}{b_U} M^2(\mu) \left[ 
   1- \left( \frac{\alpha_U(M_c)}{\alpha_U(\mu)} \right)^2 \right], \\
 && A(\mu) = -\frac{2}{b_U} \left( \sum_i C_2({\bf R}_i) \right) M(\mu)
   \left[ 1- \left( \frac{\alpha_U(M_c)}{\alpha_U(\mu)} \right) \right]. 
\end{eqnarray}

We now apply the above solutions to the minimal $SU(5)$ GUT model 
 with the particle contents as in Table \ref{su5}. 
Since the beta function coefficient of the model is $b_U=3$, we have 
 \begin{eqnarray} 
 &&  \alpha_U(M_{GUT})^{-1} =  \alpha_U(M_c)^{-1} + \frac{3}{2 \pi} 
     \ln(M_{GUT}/M_c),  \\ 
 && m^2_{\bf 10}(M_{GUT}) = \frac{12}{5} M_{1/2}^2
   \left[ 1- \left( \frac{\alpha_U(M_c)}{\alpha_U(M_{GUT})} 
   \right)^2 \right],  \\
 && m^2_{\bar{\bf 5}}(M_{GUT}) = m^2_{\bf 5}(M_{GUT}) 
   = \frac{8}{5} M_{1/2}^2 \left[ 
    1- \left( \frac{\alpha_U(M_c)}{\alpha_U(M_{GUT})} \right)^2 \right], \\
 && A_u(M_{GUT}) = -\frac{32}{5} M_{1/2}
    \left[ 1- \left( \frac{\alpha_U(M_c)}{\alpha_U(M_{GUT})} 
     \right) \right],  \\ 
 && A_d(M_{GUT}) = -\frac{28}{5} M_{1/2}
    \left[ 1- \left( \frac{\alpha_U(M_c)}{\alpha_U(M_{GUT})} \right) \right],  
\end{eqnarray}
where $M_{1/2}=M(M_{GUT})$ is the universal gaugino mass at the GUT scale. 
Note that the sfermion masses at the GUT scale are not universal, 
 but the relation between soft masses of different representation fields 
 are fixed by $C_2$.

For the simple $SO(10)$ model with the particle contents 
 in Table \ref{so10}, the beta function coefficient is $b_U=4$ 
 and we have  
\begin{eqnarray} 
 &&  \alpha_U(M_{GUT})^{-1} =  \alpha_U(M_c)^{-1} 
 + \frac{2}{\pi} \ln(M_{GUT}/M_c),  \\ 
 && m^2_{\bf 16}(M_{GUT}) = \frac{45}{16} M_{1/2}^2 
    \left[ 1- \left( \frac{\alpha_U(M_c)}{\alpha_U(M_{GUT})} 
    \right)^2 \right],  \\
 && m^2_{\bf 10}(M_{GUT}) = \frac{9}{4} M_{1/2}^2   \left[ 
    1- \left( \frac{\alpha_U(M_c)}{\alpha_U(M_{GUT})} \right)^2 \right], \\
 && A(M_{GUT}) = -\frac{63}{8} M_{1/2} 
    \left[ 1- \left( \frac{\alpha_U(M_c)}{\alpha_U(M_{GUT})} 
    \right) \right].  
\end{eqnarray}
In the $SO(10)$ model, the MSSM sfermion masses 
 are universal at the GUT scale. 

%%%%%%%%%%%%%%%%%%%%%%%%%%%%%%%%%%%%%%%%%%%%%%%
\begin{figure}[h]
\begin{center}
\includegraphics[scale=0.68]{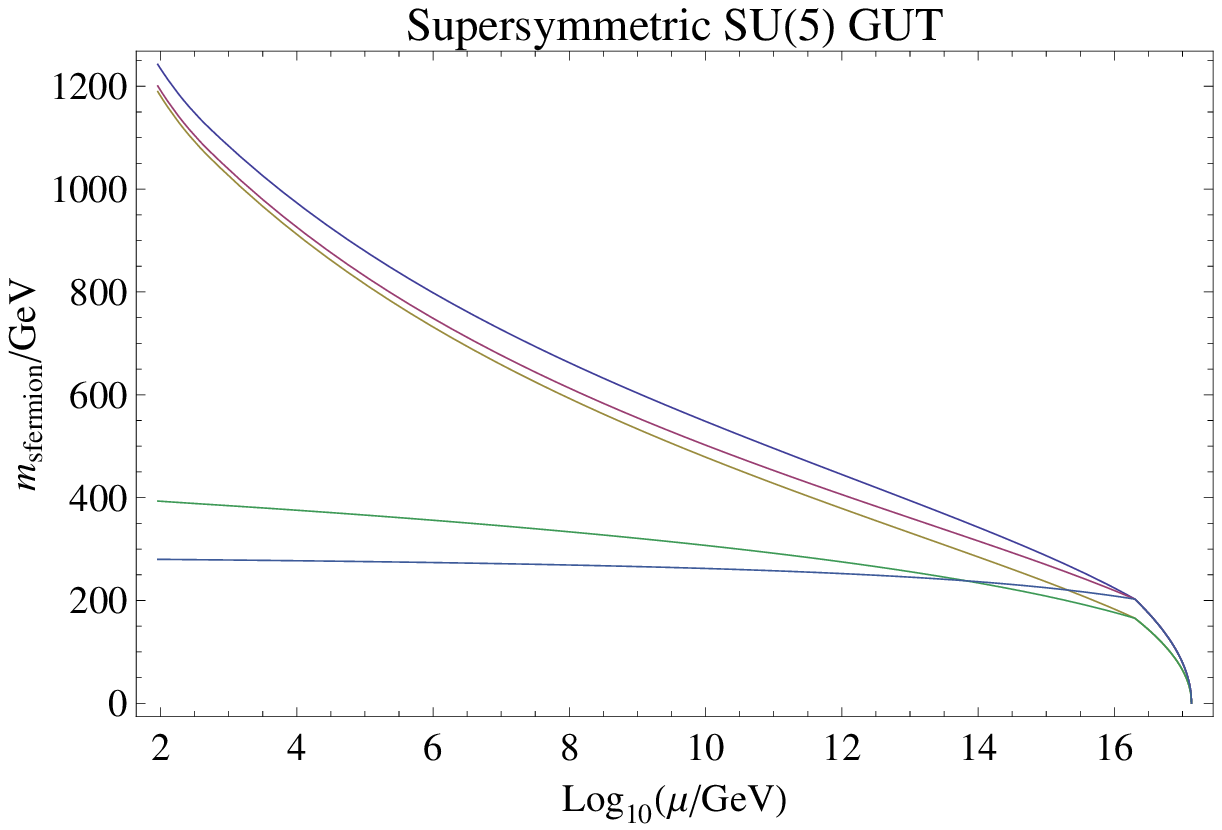}
\includegraphics[scale=0.68]{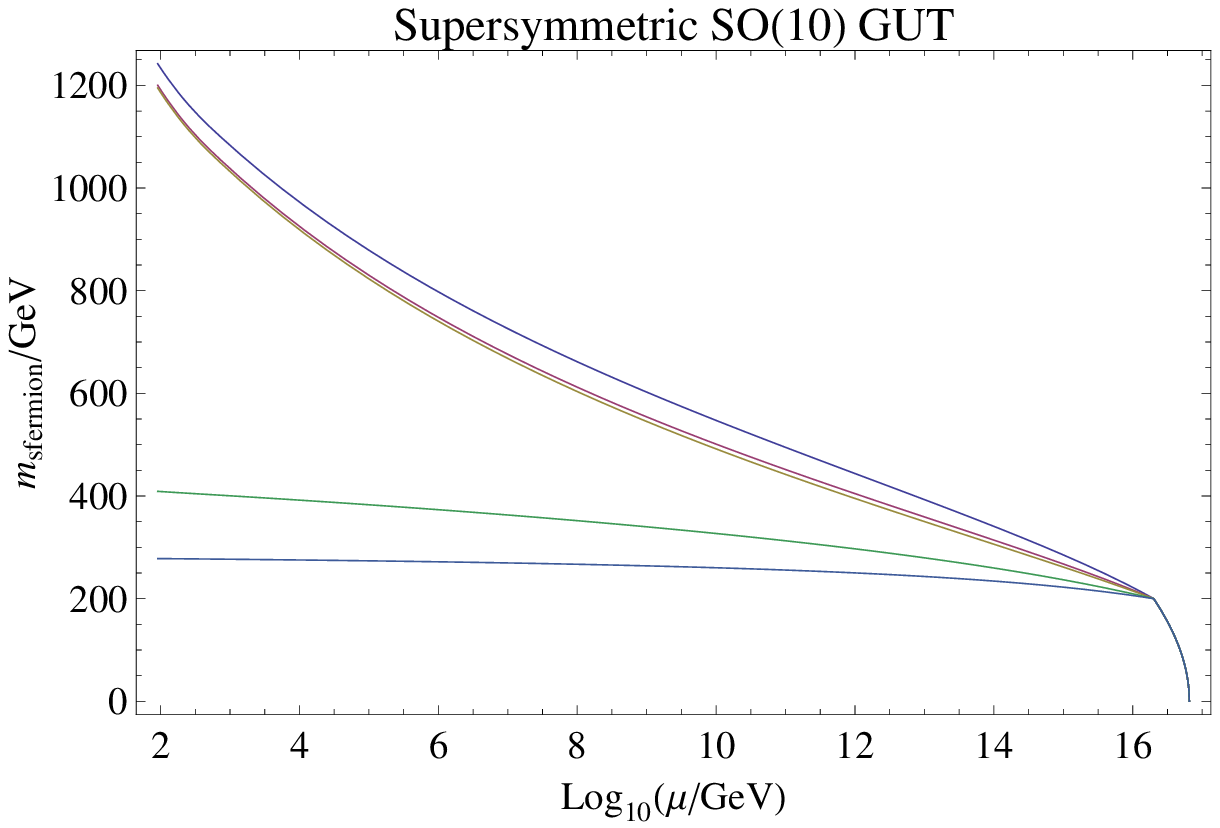}
\caption{
RGE evolution of the first two generation sfermion soft masses 
 ($m_{\tilde{Q}}$, $m_{\tilde{U}^c}$, $m_{\tilde{D}^c}$, 
  $m_{\tilde{L}}$ and $m_{\tilde{E}^c}$ from top to bottom) 
 with $\tan \beta = 30$, $\mu >0$ and $M_{1/2} = 500$ GeV 
 for the $SU(5)$ and $SO(10)$ models, respectively. 
} 
\label{running}
\end{center}
\end{figure}
%%%%%%%%%%%%%%%%%%%%%%%%%%%%%%%%%%%%%%%%%%%%%%%

For numerical calculation, we have only two free parameters, 
 $M_G$ and $M_c$, with fixed $\tan \beta$ 
 and the sign of the $\mu$-parameter. 
In MSSM RGE analysis below $M_{GUT}$, 
 we choose $M_{1/2}$ as a free parameter and 
 the other soft parameters are fixed once $M_c$ fixed. 
In order to compare sparticle spectrum in the two GUT models, 
 it is necessary to fix a common base for them. 
We choose the values of free parameters in such a way that 
 two models give the same neutralino LSP mass. 
In the gaugino mediations, neutralino LSP is binolike, 
 so that the same $M_{1/2}$ inputs for two models 
 give (almost) the same masses for neutralino LSP. 
The compactification scale $M_c$ is still left as 
 a free parameter, whose degree of freedom is used  
 to fix another sparticle mass. 
Here we impose a cosmological constraint that 
 the relic abundance of neutralino LSP is consistent 
 with the (cold) dark matter abundance measured 
 by the WMAP \cite{WMAP}: 
\begin{eqnarray}
 \Omega_{CMD} h^2 = 0.1131 \pm 0.0034. 
\label{relic}
\end{eqnarray}  
This WMAP constraint dramatically reduces 
 the viable parameter space of the models 
 as in the constrained MSSM \cite{DMconstraint}.
For a given $\tan \beta$ and a fixed $M_{1/2}$, 
 the compactification scale is completely fixed 
 by this cosmological constraint. 
As we will see, the right relic abundance is achieved 
 by the neutralino co-annihilations with the next-to-LSP 
 (mostly right-handed) stau almost degenerated with the LSP. 
For the two GUT models, the resultant next-to-LSP stau masses 
 are found to be almost the same.

The RGE evolutions of the first two generations of squarks and sleptons 
 are demonstrated in the case of $\tan \beta = 30$, $\mu >0$, and 
 $M_{1/2} = 500$ GeV for the $SU(5)$ and $SO(10)$ models 
 in Figure \ref{running}.  
The compactification scales $M_c$ for the two models  
 are fixed to give the correct neutralino relic abundance:  
 $M_c = 1.36 \times 10^{17}$ GeV and $6.53 \times 10^{16}$ GeV 
 for the $SU(5)$ and $SO(10)$ models, respectively. 
Here we can see characteristic features of running sfermion masses 
 for the two GUT models, namely, 
 sfermion masses are unified at two points in the $SU(5)$ model, 
 on the other hand, one-point unification in the $SO(10)$ model. 
The cosmological constraint requires the next-to-LSP stau, 
 which is mostly the right-handed stau, is almost degenerated 
 with the neutralino LSP, and we find 
 $m^{SU(5)}_{\bf 10} \approx m^{SO(10)}_{\bf 16}$ at the GUT scale. 
However, there is a sizable mass splitting 
 between $m^{SU(5)}_{\bf 5}$ and $m^{SO(10)}_{\bf 16}$. 
This is the key to distinguish the two GUT models. 
In terms of sparticles in the MSSM, 
 the difference appears in masses of down-type squarks 
 and the left-handed sleptons.

%%%%%%%%%%%%%%%%%%%%%%%%%%%%%%%%%%%%%%%%%%%%%%%%%%
\begin{figure}[h]
\begin{center}
\includegraphics[scale=0.7]{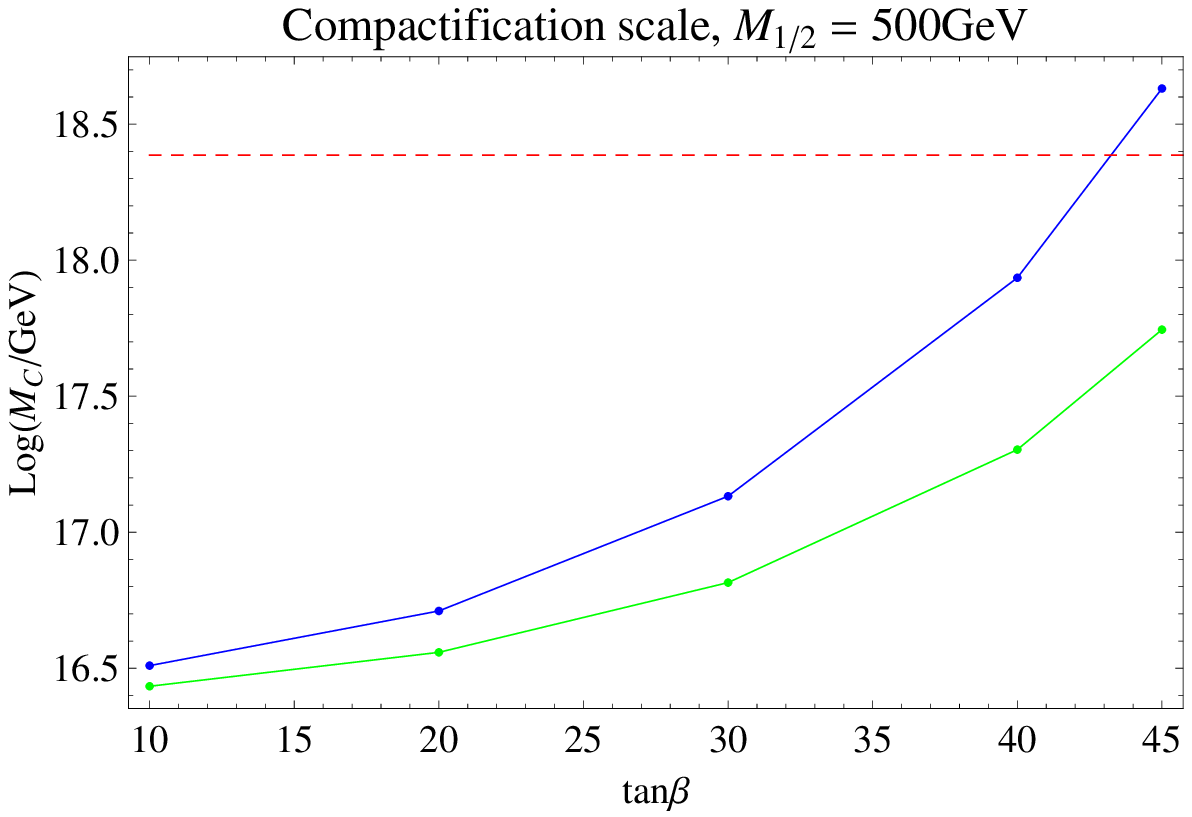}
\includegraphics[scale=0.7]{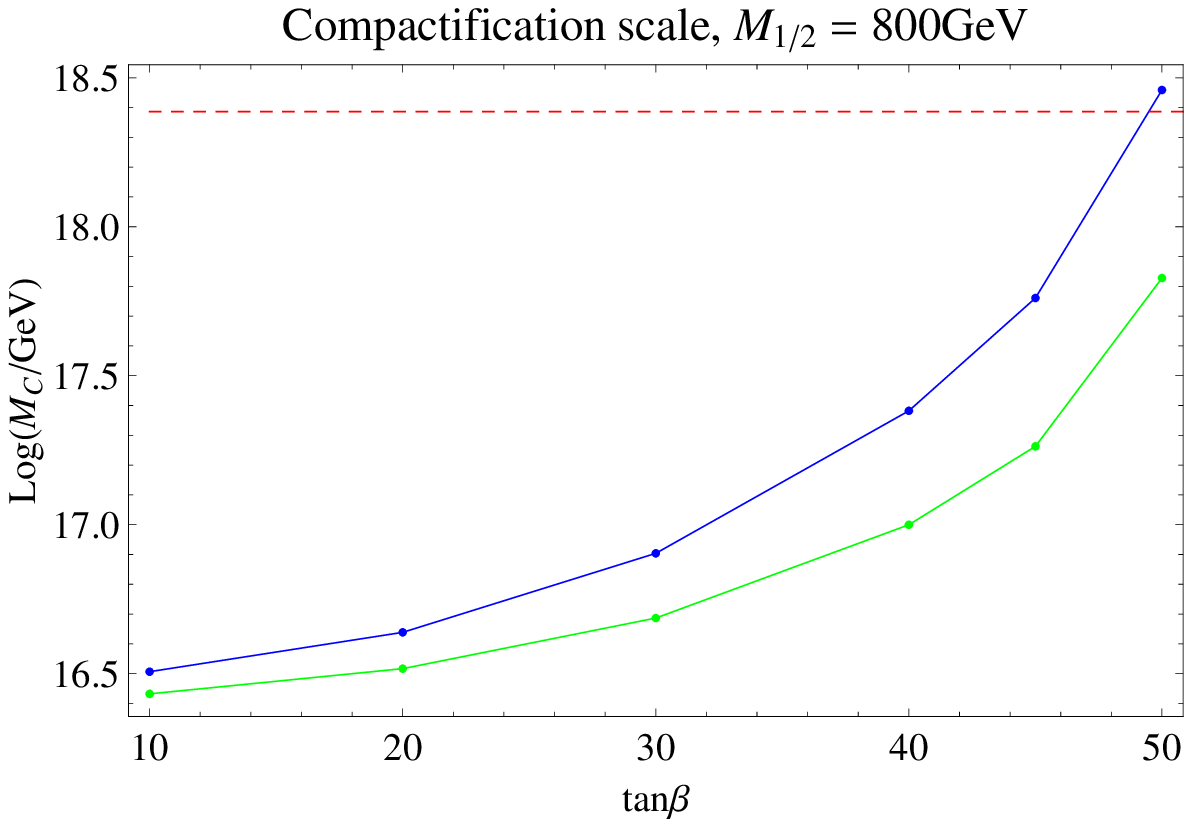}
\caption{
Compactification scale as a function of $\tan \beta$ 
 in the case $M_{1/2} = 500$ GeV and 800 GeV. 
In each plot, the upper (blue) and lower (green) solid lines 
 correspond to the $SU(5)$ and $SO(10)$ models, respectively. 
The horizontal dashed (red) line indicates the theoretical 
 constraint (\ref{Planck}).
} 
\label{compact}
\end{center}
\end{figure}
%%%%%%%%%%%%%%%%%%%%%%%%%%%%%%%%%%%%%%%%%%%%%%%%%%

In our numerical analysis, we employ 
 the SOFTSUSY 3.1.4 package \cite{softsusy} 
 to solve the MSSM RGEs and produce mass spectrum. 
While running this program, we always set $sign(\mu) = +1$, 
 for simplicity. 
The relic abundance of the neutralino dark matter 
 is calculated by using the micrOMEGAs 2.4 \cite{micromega} 
 with the output of SOFTSUSY in the SLHA format \cite{slha}. 
In addition to the cosmological constraint, 
 we also take into account other phenomenological constraints 
 such as the lower bound on Higgs boson mass \cite{Higgs}:
\begin{eqnarray} 
m_{h} \geq 114.4 \; \rm GeV, 
\label{mh} 
\end{eqnarray}
 the constraints on the branching ratios of 
 $b \rightarrow s \gamma$, $B_{s} \rightarrow \mu^{+} \mu^{-} $ 
 and the muon anomalous magnetic moment 
 $\Delta a_{\mu} = g_{\mu} - 2$: 
\begin{eqnarray}
& 2.85 \times 10^{-4} \leq BR(b \rightarrow s + \gamma) \leq 4.24
\times 10^{-4} \; (2 \sigma ) & \hspace{1cm} \cite{bsgamma} 
\label{bsg}, \\
&  BR(B_{s} \rightarrow \mu^{+} \mu^{-} ) < 5.8 \times 10^{-8}  &
\hspace{1cm} \cite{bsmumu},
 \label{bsmm} \\
& 3.4 \times 10^{-10} \leq \Delta a_{\mu} \leq 55.6 \times 10^{-10} \;
(3 \sigma )  & \hspace{1cm} \cite{delta} \label{delta}. 
\end{eqnarray}

We examine two typical values of $M_{1/2}=500$ and 800 GeV 
 for a variety of $\tan \beta =10$, 20, 30, 40, 45, and 50. 
The mass spectra of the two models are shown in Table \ref{m500} 
 for the case of $M_{1/2} = 500$ GeV and in Table \ref{m800} 
 for the case of $M_{1/2} = 800$ GeV. 
In the tables, we also list the values of the compactification scale
 $M_c$ chosen to reproduce the observed dark matter abundance, 
 the branching ratios of $b \rightarrow s \gamma$ and 
 $B_s \rightarrow \mu^+ \mu^-$, and the anomalous magnetic moment 
 of muon $\Delta a_\mu$.

Using the data in Tables \ref{m500} and \ref{m800}, 
 we plot the compactification scale 
 as a function of $\tan \beta$ for $M_{1/2}=500$ and 800 GeV, 
 respectively, in Figure \ref{compact}.  
The upper (blue) and lower (green) solid lines indicate 
 the $SU(5)$ and $SO(10)$ models, respectively. 
The horizontal dashed (red) line corresponds to 
 the upper bound on the compactification scale (\ref{Planck}).
These figures show that the theoretical constraint (\ref{Planck}) 
 rules out a large $\tan \beta$ region for the $SU(5)$ model. 
We find the upper bounds 
 $\tan \beta \lesssim 43$ for $M_{1/2} = 500$ GeV 
 and $\tan \beta \lesssim 49$ for $M_{1/2} = 800$ GeV. 
Comparing the two plots in Figure \ref{compact}, 
 we see that the bound on $\tan \beta$ 
 becomes more severe for smaller $M_{1/2}$ inputs.

\begin{figure}[h]
\begin{center}
\includegraphics[scale=0.7]{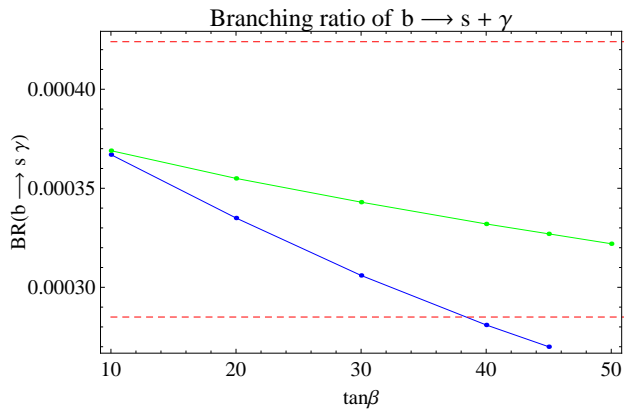}
\caption{
 $BR(b \rightarrow s \gamma)$ as a function 
 of $\tan \beta$ for $M_{1/2}=500$ and 800 GeV. 
The lower (blue) and upper (green) solid lines 
 correspond to $M_{1/2}= 500$ GeV and 800 GeV, respectively. 
The horizontal dashed (red) lines indicate 
 the upper and lower bounds 
 of the branching ratio (\ref{bsg}).}
\label{bsgplot}
\end{center}
\end{figure}

For the sparticle spectra presented 
 in Tables \ref{m500} and \ref{m800}, 
 phenomenological constraints of 
 (\ref{relic}), (\ref{mh}), (\ref{bsmm}) and (\ref{delta}) 
 are all satisfied. 
However, the predicted branching ratio 
  $BR(b \rightarrow s \gamma)$ can be too small 
 to satisfy the experimental bound (\ref{bsg})
 for a large $\tan \beta$. 
In Figure \ref{bsgplot}, we show the values of 
 $BR(b \rightarrow s \gamma)$ for all the samples 
 in Table \ref{m500} and \ref{m800}, 
 along with the experimental allowed region 
 between two dashed (red) lines. 
We can see that for the case with $M_{1/2} = 500$ GeV, 
 there is  an upper bound on $\tan \beta \lesssim 38$.   
In general, for a smaller $M_{1/2}$ input,  
 we will find a more severe bound on $\tan \beta$.

%%%%%%%%%%%%%%%%%%%%%%%%%%%%%%%%%%%%%%%%%%%%%%%%%
%%%%%%%%%%%%%%%%%%%%%%%%%%%%%%%%%%%%%%%%%%%%%%%%%

\begin{figure}[h]
\begin{center}
\includegraphics[scale=0.7]{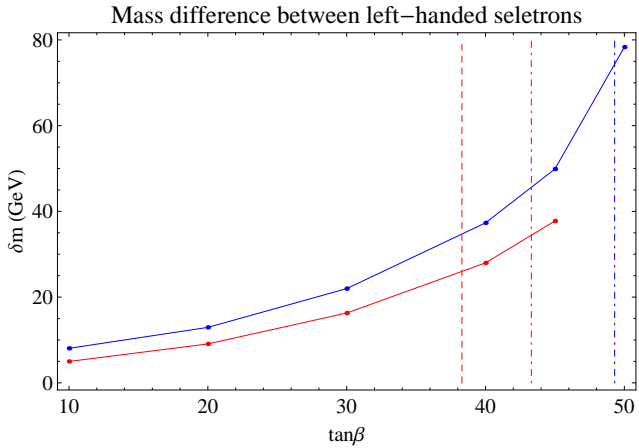}
\caption{
Mass difference $\delta m = m^{SO(10)} - m^{SU(5)}$ 
 between left-handed selectrons/smuons of the two models 
 is plotted as a function of $\tan \beta$ 
 for $M_{1/2}=500$ and 800 GeV. 
The lower (red) and upper (blue) solid lines correspond 
 to Table \ref{m500} with $M_G = 500$ GeV and Table \ref{m800} 
 with $M_G = 800$ GeV, respectively. 
The dashed line is the upper bound on $\tan \beta$ 
 from the $b \rightarrow s \gamma$ constraint. 
The dot-dashed lines indicate the upper bounds 
 on $\tan \beta$ by the theoretical constraint $M_c < M_P$. 
The right vertical bound applies to the case with $M_{1/2}=800$ GeV, 
 while two left vertical lines to the case with $M_{1/2}=500$ GeV. 
}
\label{difference}
\end{center}
\end{figure}

Taking into account all theoretical and phenomenological bounds, 
 we compare the mass difference between the two GUT models. 
As mentioned before, in Tables \ref{m500} and \ref{m800} 
 we see relatively large mass differences 
 in left-handed slepton sector and right-handed down-type squark sector. 
This effect is not so clear in the third-generation squark masses 
 because of the large Yukawa contributions. 
Figure \ref{difference} shows the mass difference 
 $\delta m = m^{SO(10)} - m^{SU(5)}$ 
 between left-handed selectrons/smuons of the two models 
 as a function of $\tan \beta$ for $M_{1/2}=500$ GeV (lower solid line) 
 and 800 GeV (upper solid line). 
As we have discussed above, the upper bound on $M_c$ and 
 the constraint from sparticle contributions 
 to the $b \rightarrow s \gamma$ process 
 provide us the upper bound on $\tan \beta$. 
The dashed vertical line and the left dot-dashed line 
 correspond to the upper bound on $\tan \beta$ from 
 $ BR(b \to s \gamma) $ and $M_c \leq M_P$, 
 respectively, applied to the case with 
 $M_{1/2}=500$ GeV (lower solid line). 
The right dot-dashed line is the upper bound 
 from $ BR(b \to s \gamma)$ for the case 
 with $M_{1/2}=800$ GeV (upper solid line).  
Depending on values of $\tan \beta$, 
 the mass differences for $M_{1/2}=500$ GeV varies 
 $\delta m =5-25$ GeV, while $\delta m =7-75$ GeV 
 for $M_{1/2}=800$ GeV. 
These mass differences can be sufficiently large compared 
 to expected errors in measurements of sparticle masses 
 at the LHC and the ILC \cite{BBPT}.

%%%%%%%%%%%%%%%%%%%%%%%%%%%%%%
\section{Conclusion}
%%%%%%%%%%%%%%%%%%%%%%%%%%%%%%

In the context of the gaugino mediation scenario, 
 we have investigated supersymmetric grand unified theories. 
The gaugino mediation scenario, once applied to a GUT model, 
 is highly predictive and all sparticle masses are 
 determined by only two inputs, the unified gaugino mass 
 and the compactification scale, 
 with a given $\tan \beta $ and the sign of the $\mu$-parameter. 
When we choose a particular GUT model with fixed particle contents,  
  the relation among sparticle masses at the GUT scale 
  is determined by the group theoretical factors, 
  the Dynkin index and the quadratic Casimir, 
  associated with the representation of fields. 
Therefore, the difference of GUT models is reflected 
 in sparticle mass spectrum at low energies. 
Taking two GUT models, the minimal $SU(5)$ GUT and 
 simple $SO(10)$ GUT models as examples, 
 we have analyzed sparticle mass spectra 
 together with theoretical and phenomenological constraints 
 and compared resultant sparticle masses in the two models. 
Because of the difference in unification of quarks and leptons 
 into representations under the GUT gauge groups, 
 a significant difference among sparticle masses appears 
 in the left-handed slepton and right-handed down-type squark sectors. 
Fixing the input parameters in each model so as to give 
 the same neutralino mass and to reproduce the observed 
 neutralino dark matter relic abundance, 
 we have found sizable differences in sparticle mass spectra 
 in two models, which can be identified in the LHC and the ILC. 
Although we have considered only two GUT models,
 our strategy is general, 
 and we conclude that precise measurements of sparticle 
 mass spectrum can be a probe to discriminate 
 various supersymmetric unification scenarios.

Finally, we give a comment on the upper bound of the compactification scale $M_c \leq M_P$ (Eq. (\ref{Planck})).
For a large $\tan \beta$, we need to raise $M_c$ 
 close to $M_P$ in order to make neutralino the LSP and to obtain the correct relic abundance of neutralino dark matter.
In this case, the sequestering effect becomes weaker and 
 the boundary conditions set as $m_0(M_c)=0$ and $A_0(M_c)=0$ 
 in our analysis will be no longer valid. 
Despite the fact that the tree level contributions to $m_0 (M_c)$ and $A_0(M_c)$  
 remain zero, their nonzero values can be
 induced by loop effects of bulk fields such as the bulk gauge  
 and the bulk supergravity multiplets. 
For example, the contributions to $m_0^2$ have been explicitly calculated as 
\bea 
 \Delta (m_0^2)_{gauge} = \frac{\alpha_U(M_c)}{4 \pi}{M_G^2}
\eea  
 for the bulk gauge contribution \cite{gMSB}, 
 while for the bulk supergravity contribution \cite{5DSUGRA}, 
\bea 
 \Delta (m_0^2)_{sugra} = - \frac{1}{16 \pi^2} m_{3/2}^2 \left(\frac{M_c}{M_P}\right)^2
\eea  
 with $m_{3/2}$ being gravitino mass. 
In the gaugino mediation scenario, we have a relation 
 $m_{3/2} \simeq M_G (M_P/M_c)^{1/3}$ \cite{spectrumgMSB} 
 and thus, the supergravity contributions is rewritten as 
\bea 
 \Delta (m_0^2)_{sugra} = - \frac{1}{16 \pi^2} M_G^2 \left(\frac{M_c}{M_P}\right)^{4/3}. 
\eea  
Note that although there is no volume suppression effect by $M_c/M_P$ 
 when $M_c \simeq M_P$, these contributions are still loop-suppressed. 
For $M_c \simeq M_P$, we have estimated that the nonzero $m_0(M_c)$ 
 causes about 1\% changes in resultant sparticle mass spectrum. These loop corrections are negligible.

\section*{Acknowledgment}

H.M.T. would like to thank the organizers of 
 the KEK-Vietnam visiting program, especially, 
 Yoshimasa Kurihara, for their hospitality and supports 
 during his visit. 
The work of N.O. is supported in part by DOEGrantNo. DE-FG02-10ER41714.

%%%%%%%%%%%%%%%%%%%%%%%%%%%%%%%%%

\newpage

% Table of results:

\begin{landscape}

\begin{table}
\caption{Mass spectra and constraints for the two SUSY GUT models in gaugino mediation with $M_{1/2} = 500$ GeV} \label{m500}
\scalebox{0.61}[0.65]{
\begin{math}
\begin{array}{|c|c|c|c|c|c|c|c|c|c|c|}
\hline  & SU(5) & SO(10) & SU(5) & SO(10) & SU(5) & SO(10) & SU(5) & SO(10) & SU(5) & SO(10) \\ 
\hline \tan \beta & \multicolumn{2}{|c|}{10}  & \multicolumn{2}{|c|}{20} & \multicolumn{2}{|c|}{30} & \multicolumn{2}{|c|}{40}  & \multicolumn{2}{|c|}{45}  \\ 
\hline h_0 & 115 & 115 & 116 & 116 & 117 & 117 & 117 & 117 & 117 & 117 \\ 
 H_0 & 720 & 720 & 684 & 683 & 639 & 636 & 582 & 573 & 551 & 535 \\ 
 A_0 & 719 & 720 & 684 & 683 & 639 & 636 & 583 & 573 & 551 & 535 \\
 H^\pm & 724 & 724 & 689 & 688 & 645 & 642 & 588 & 579 & 557 & 541 \\  
\hline \tilde{g} & 1146 & 1146 & 1147 & 1147 & 1148 & 1148 & 1151 & 1151 & 1153 & 1153 \\ 
 {\tilde{\chi}^0}_{1,2,3,4} & 204, 387, 649, 662 & 204, 387, 649, 662 & 205, 389, 652, 663 & 205, 389, 652, 663 & 206, 391, 666, 676 & 206, 391, 666, 676 & 206, 393, 694, 703 & 206, 393, 693, 702 & 207, 395, 717, 725 & 207, 395, 717, 725 \\ 
 {\tilde{\chi}^{\pm}}_{1,2} & 387, 662 & 387, 662 & 389, 663 & 389,663 & 391, 676 & 391, 676 & 393, 703 & 393, 702 & 395, 725 & 395, 725 \\ 
\hline \tilde{d},\tilde{s}_{R,L} & 1007, 1053 & 1009, 1053 & 1010, 1058 & 1013, 1058 & 1017, 1068 & 1023, 1067 & 1028, 1084 & 1040, 1083 & 1037, 1097 & 1054, 1096 \\ 
 \tilde{u},\tilde{c}_{R,L} & 1012, 1051 & 1012, 1050 & 1017, 1055 & 1017, 1055 & 1027, 1065 & 1026, 1064 & 1044, 1081 & 1043, 1080 & 1058, 1094 & 1057, 1094 \\ 
 \tilde{b}_{1,2} & 963, 1004 & 963, 1005 & 954, 998 & 955, 1000 & 940, 990 & 941, 994 & 921, 985 & 924, 989 & 910, 984 & 916, 989 \\ 
 \tilde{t}_{1,2} & 801, 1010 & 801, 1010 & 805, 1006 & 805, 1006 & 808, 1003 & 807, 1002 & 812, 1002 & 810, 1000 & 814, 1003 & 812, 1000 \\ 
\hline \tilde{\nu}_{e,\mu,\tau} & 341, 341, 340 & 346, 346, 345 & 350, 350, 346 & 360, 360, 355 & 369, 369, 357 & 386, 386, 373 & 400, 400, 374 & 428, 428, 402 & 422, 422, 386 & 461, 461, 424 \\ 
 \tilde{e},\tilde{\mu}_{R,L} & 219, 350 & 219, 355 & 241, 359 & 240, 368 & 280, 378 & 278, 394 & 337, 408 & 335, 436 & 377, 430 & 376, 468 \\ 
 \tilde{\tau}_{1,2} & 211, 351 & 211, 356 & 211, 364 & 211, 372 & 214, 386 & 214, 399 & 219, 417 & 218, 436 & 222, 436 & 221, 461 \\ 
\hline M_c & 3.23 \times 10^{16} & 2.71 \times 10^{16} & 5.14 \times 10^{16} & 3.62 \times 10^{16} & 1.36 \times 10^{17} & 6.53 \times 10^{16} & 8.62 \times 10^{17} & 2.01 \times 10^{17} & 4.28 \times 10^{18} & 5.56 \times 10^{17} \\ 
 BR(b \rightarrow s \gamma) & 3.67 \times 10^{-4} & 3.67 \times 10^{-4} & 3.35 \times 10^{-4} & 3.35 \times 10^{-4} & 3.06 \times 10^{-4} & 3.06 \times 10^{-4} & 2.81 \times 10^{-4} & 2.81 \times 10^{-4} & 2.70 \times 10^{-4} & 2.69 \times 10^{-4} \\ 
 BR(B_s \rightarrow \mu^+ \mu^-) & 3.15 \times 10^{-9} & 3.15 \times 10^{-9} & 3.59 \times 10^{-9} & 3.59 \times 10^{-9} & 5.83 \times 10^{-9} & 5.87 \times 10^{-9} & 1.67 \times 10^{-8} & 1.75 \times 10^{-8} & 3.42 \times 10^{-8} & 3.79 \times 10^{-8} \\ 
 \Delta a_{\mu} & 9.28 \times 10^{-10} & 9.13 \times 10^{-10} & 1.74 \times 10^{-9} & 1.69 \times 10^{-9} & 2.35 \times 10^{-9} & 2.25 \times 10^{-9} & 2.71 \times 10^{-9} & 2.53 \times 10^{-9} & 2.78 \times 10^{-9} & 2.55 \times 10^{-9} \\ 
\hline \Omega h^2 & \multicolumn{2}{|c|}{0.113}  & \multicolumn{2}{|c|}{0.113} & \multicolumn{2}{|c|}{0.113} & \multicolumn{2}{|c|}{0.113} & \multicolumn{2}{|c|}{0.113} \\ 
\hline 
\end{array}
\end{math}}
\end{table}

\begin{table}
\caption{Mass spectra and constraints for the two SUSY GUT models in gaugino mediation with $M_{1/2} = 800$ GeV} \label{m800}
\scalebox{0.485}[0.63]{
\begin{math} 
\begin{array}{|c|c|c|c|c|c|c|c|c|c|c|c|c|}
\hline  &  SU(5) &  SO(10) & SU(5) &  SO(10) &  SU(5) &  SO(10) & SU(5) & SO(10) & SU(5) & SO(10) & SU(5) & SO(10) \\
\hline  \tan \beta & \multicolumn{2}{|c|}{10} & \multicolumn{2}{|c|}{20} & \multicolumn{2}{|c|}{30} & \multicolumn{2}{|c|}{40} & \multicolumn{2}{|c|}{45} & \multicolumn{2}{|c|}{50} \\
\hline h_0 & 119 & 119 & 119 &  119 & 119 &  119 & 119 & 119 & 119 & 119 & 119 &  119 \\
       H_0 & 1106 & 1106 & 1049 & 1048 & 975 & 972 & 877 & 868 & 819 & 805 & 762 & 737 \\
       A_0 & 1106 & 1106 & 1049 & 1048 & 975 & 972 & 877 & 869 & 820 & 805 & 762 & 737 \\
     H^\pm & 1109 & 1109 & 1052 & 1051 & 978 & 976 & 881 & 873 & 824 & 809 & 767 & 742 \\
\hline \tilde{g} & 1770 & 1771 & 1771 & 1771 & 1772 & 1772 & 1775 & 1775 & 1777 & 1778 & 1780 & 1783 \\
{\tilde{\chi}^0}_{1,2,3,4} & 335, 634, 983, 992 & 335, 635, 983, 993 & 336, 636, 982, 990 & 336, 636, 982, 991 & 337, 638, 995, 1003 & 337, 638, 996, 1004 & 338, 640, 1022, 1029 & 338, 641, 1024, 1031 & 338, 642, 1043, 1049 & 338, 643, 1048, 1054 & 339, 644, 1081, 1087 & 340, 646, 1099, 1104 \\
{\tilde{\chi}^\pm}_{1,2} & 635, 992 & 635, 992 & 636, 991 & 637, 991 & 638, 1003 & 639, 1004 & 640, 1029 &  641, 1031 &  642, 1050 &  643, 1055 &  644, 1087 &  646, 1105 \\
\hline \tilde{d},\tilde{s}_{R,L} & 1544, 1619 & 1547, 1619 & 1547, 1625 & 1552, 1625 & 1554, 1635 & 1563, 1635 & 1567, 1652 & 1582, 1653 & 1576, 1665 & 1597, 1667 & 1592, 1687 & 1626, 1695 \\
\tilde{u},\tilde{c}_{R,L} & 1553, 1618 & 1553, 1618 & 1559, 1623 & 1559, 1623 & 1569, 1633 & 1569, 1633 & 1588, 1650 & 1588, 1651 & 1601, 1663 & 1603, 1665 & 1624, 1686 & 1633, 1694 \\
\tilde{b}_{1,2} & 1485, 1537 & 1485, 1540 & 1474, 1523 & 1474, 1527 & 1454, 1505 & 1456, 1510 & 1427, 1489 & 1432, 1494 & 1411, 1483 & 1419, 1489 & 1396, 1482 & 1411, 1491 \\
\tilde{t}_{1,2} & 1254, 1517 & 1254, 1517 & 1259, 1510 & 1259, 1510 & 1263, 1502 & 1263, 1501 & 1269, 1495 & 1269, 1493 & 1273, 1492 & 1273, 1491 & 1279, 1494 & 1279, 1493 \\
\tilde{\nu}_{e,\mu,\tau} & 546, 546, 545 & 555, 555, 553 & 557, 557, 549 & 570, 570, 562 & 576, 576, 559 & 598, 598, 581 & 608, 608, 574 & 646, 646, 611 & 631, 631, 585 & 682, 682, 633 & 669, 669, 604 & 748, 748, 677 \\
\tilde{e},\tilde{\mu}_{R,L} & 345, 552 & 345, 560 & 368, 562 & 369, 575 & 411, 582 & 411, 604 & 475, 614 & 478, 651 & 518, 636 & 525, 686 & 585, 674 & 609, 752 \\
\tilde{\tau}_{1,2} & 337, 552 & 337, 560 & 338, 561 & 338, 574 & 341, 578 & 340, 597 & 346, 603 & 346, 634 & 351, 619 & 350, 660 & 367, 645 & 370, 706 \\
\hline M_c & 3.21 \times 10^{16} & 2.70 \times 10^{16} & 4.35 \times 10^{16} & 3.29 \times 10^{16} & 8.01 \times 10^{16} & 4.86 \times 10^{16} & 2.41 \times 10^{17} & 9.99 \times 10^{16} & 5.76 \times 10^{17} & 1.83 \times 10^{17} & 2.88 \times 10^{18} & 6.73 \times 10^{17} \\
BR(b \rightarrow s \gamma) &   3.69 \times 10^{-4} &   3.69 \times 10^{-4} &   3.55 \times 10^{-4} &   3.55 \times 10^{-4} &   3.43 \times 10^{-4} &   3.43 \times 10^{-4} &   3.32 \times 10^{-4} &   3.32 \times 10^{-4} &   3.26 \times 10^{-4} &   3.27 \times 10^{-4} &   3.22 \times 10^{-4} &   3.22 \times 10^{-4} \\
BR(B_s \rightarrow \mu^+ \mu^-) &   3.13 \times 10^{-9} &   3.13 \times 10^{-9} &   3.28 \times 10^{-9} &   3.29 \times 10^{-9} &   4.01 \times 10^{-9} &   4.02 \times 10^{-9} &   6.89 \times 10^{-9} &   7.02 \times 10^{-9} &   1.10 \times 10^{-8} &   1.16 \times 10^{-8} &   2.03 \times 10^{-8} &   2.31 \times 10^{-8} \\
\Delta a_\mu &   3.61 \times 10^{-10} &   3.55 \times 10^{-10} &   6.91 \times 10^{-10} &   6.74 \times 10^{-10} &   9.65 \times 10^{-10} &   9.26 \times 10^{-10} &   1.16 \times 10^{-9} &   1.09 \times 10^{-9} &   1.23 \times 10^{-9} &   1.13 \times 10^{-9} &   1.24 \times 10^{-9} &   1.09 \times 10^{-9} \\
\hline \Omega h^2 &  \multicolumn{2}{|c|}{0.113} &  \multicolumn{2}{|c|}{0.113} & \multicolumn{2}{|c|}{0.113} & \multicolumn{2}{|c|}{0.113} & \multicolumn{2}{|c|}{0.113} &  \multicolumn{2}{|c|}{0.113} \\
\hline
\end{array}
\end{math}}
\end{table} 

\end{landscape}

\end{document}